# Some Enigmatic Aspects of the Early Universe II

C Sivaram, Kenath Arun and Venkata Manohara Reddy A

Indian Institute of Astrophysics, Bangalore

**Abstract:** In a recent paper it was suggested that inclusion of mutual gravitational interactions among the collapsing particles can avert a singularity and give finite value for various physical quantities. In this paper we extend this idea further by the inclusion of charge and spin to the system. We have also discussed other possible scenarios by which the singular state can be averted, including a temperature dependant gravitational constant. Also possible modifications in the Einstein-Hilbert action have been discussed, which can again lead to a finite maximal curvature hence avoiding a singularity.

In a recent paper[1] we had discussed a possible scenario in the early universe or alternately in the case of the universe recollapsing to a singularity, where the inclusion of the mutual gravitational interactions among the collapsing particles can avert a singularity and give finite values for various physical quantities like entropy, density, etc. The singularity theorems propose that the conditions for a singularity to be inevitable include the global positivity of the energy density (that is, $\rho + 3P \geq 0$) and also the non-existence of the positive $\Lambda$ term.

In the preceding paper[1] we have discussed the situation where the total energy is zero and we had a positive $\Lambda$ term. In ref. [1] the effects of charge and spin were not considered. Including these effects, the total mass of the system can be written as:[2]

$$M_{Total} = M_0 - \frac{GM_{Total}^2}{2Rc^2} + \frac{e^2}{2Rc^2} + \frac{J}{2Rc} \qquad \ldots (1)$$

Where, $M_0$ is the bare mass, $R$ and $e$ are the size and charge associated with the system and $J$ is the angular momentum. This is a simple quadratic equation and in the limit of $R \to 0$ and $M_0 = 0$ the solution works out to be:

$$M_{Total} = \left[\frac{e^2}{G} + \frac{Jc}{2G}\right]^{1/2} \qquad \ldots (2)$$

For an electrically neutral system, we have:

$$M_{Total} = \left[\frac{Jc}{2G}\right]^{1/2} \qquad \ldots (3)$$

For angular momentum, $J = \hbar$, we have:

$$M_{Total} = \left[\frac{\hbar c}{2G}\right]^{1/2} \qquad \ldots (4)$$

Which is the Planck mass $(\sim 10^{-5} g)$.

For the universe, the relation becomes:[1] $J \sim 10^{120} \hbar \Rightarrow M_{Total} = 10^{60} M_{Pl}$

The situation in general relativity is very similar, the net mass there being the ADM mass; given by an equation similar to that of equation (1).[3]

In the case of a galaxy, $J \sim 10^{100} \hbar \Rightarrow M_{Total} = 10^{50} M_{Pl} \sim 10^{12} M_{\Theta}$ and that for the solar system, $J \sim 10^{76} \hbar \Rightarrow M_{Total} = 10^{38} M_{Pl} \sim 10^{33} g$.[4]

For the Universe we saw that the total mass is given by $M_{Total} = 10^{60} M_{Pl}$. If it collapses to the Planck density then the size associated (with a number density of $N = 10^{60}$) can be written as: $N^{\frac{1}{3}} L_{Pl} = 10^{20} L_{Pl} \sim 10^{-13} cm$. Then expansion of the universe started out at $10^{-13} cm$ and continued through $10^{-3} cm$, keeping the angular momentum $(J \sim 10^{120} \hbar)$ constant and the density $\rho = \rho_{Pl}$.

In the inflationary expansion from $10^{-13} cm$ to $10^{-3} cm$, at constant Planck density, the mass increased by a factor of $(10^{10})^3$, so that the total mass associated with the universe at the epoch corresponding to $R = 10^{-3} cm$ is given by: $M_{Total} = 10^{90} M_{Pl}$

Further expansion from $10^{-3} cm$ to $10^{28} cm$ is determined by (the usual Robertson-Walker equation):

$$\frac{\dot{R}}{R} = \left(\frac{8\pi G \rho}{3c^2}\right)^{\frac{1}{2}} \quad \ldots (5)$$

This involves only the energy density term. The gravitational self energy and the curvature terms cancel out. This can be seen from the relation:[5]

$$\frac{GM_U^2}{R} = \Lambda c^2 M_U R^2 \quad \ldots (6)$$

The curvature corresponding to this condition is given by:[5, 6]

$$\Lambda = \frac{GM_U}{R^3 c^2} \sim 10^{66} cm^{-2} \quad \ldots (7)$$

Where, $M_U = 10^{90} M_{Pl}$ (as obtained above), $R = 10^{-3} cm$. The curvature obtained corresponds to the Planck curvature!

So from $10^{-13} cm$ to $10^{-3} cm$, the expansion was dominated by the curvature term. At $10^{-3} cm$, we have only the energy density term as given by equation (5). Initially at $R = 10^{-3} cm$, we had all the three terms.[1]

From the scale of $R = 10^{-3} cm$, the radiation domination era began. (The initial temperature being $T = T_{Pl} \approx 10^{-32} K$). During the subsequent expansion (which was just Friedman, dominated by the energy density term) $RT$ remained constant.[1] The expansion rate for radiation followed: $R(t) \propto t^{1/2}$.

The expansion became matter dominated at $T \approx 4000K$, when radiation and matter had comparable densities. The matter dominated era then took over, having a dependence $R(t) \propto t^{2/3}$.

The discussions above give a possible scenario for a non-singular origin of the universe and possibly recollapse into non-singular state as implied by the equations (5) to (7) (also see equations (9) to (12) of ref. [1]).

Another possible way of averting a singular state is to consider a temperature (energy) dependent gravitational constant with $G$ depending on $T$ as:[7, 8]

$$G = G_0\left(1 - \frac{T_{Pl}^2}{T^2}\right) \quad \ldots (8)$$

Then as the temperature tends to the Planck temperature (as energy tends to $E_{Pl}$) the coupling vanishes $(G(T) \to 0)$, implying that the curvature also tends to zero, hence avoiding the singularity.[9]

As the temperature becomes much less than the Planck temperature, $T \ll T_{Pl}$, we get back the Newtonian value for $G$, that is $G = G_0$.

An analogy of this can be seen in the case of superconductors. The magnetic field inside a superconductor is given by the relation:[10]

$$B = B_0\left(1 - \frac{T_C^2}{T^2}\right) \qquad \ldots (9)$$

Where, $T_C$ is the critical transition temperature. As the temperature reduces to less than the transition temperature, the field inside the superconductor vanishes, exhibiting perfect diamagnetism by the material.

Here similarly $G$ vanishes at $T > T_{Pl}$. This weakening of gravity above a certain critical temperature would avoid the formation of singular state in a gravitational collapse.

This argument is keeping in tune with the concept of asymptotic freedom which characterise other theories like the strong interaction. The coupling vanishes as energies and temperature tends to very large values.[7, 9]

Along with vanishing gravitational constant we can have curvature and other curvature invariants going to zero as the temperature increases beyond the Planck temperature. We have the relation $R = \kappa T'$ ($T'$ being the trace of the energy momentum tensor), where the curvature can have the same temperature dependence as the gravitational constant as:[7, 8]

$$\kappa = \kappa_0\left(1 - \frac{T_{Pl}^2}{T^2}\right) \qquad \ldots (10)$$

The rate of expansion from equation (5) can be written as:

$$\frac{\dot{T}}{T} = -\frac{\dot{R}}{R} = \left(\frac{8\pi G a}{3c^2}\right)^{1/2} T^2 \qquad \ldots (11)$$

Where, the energy density can be written as: $\rho = aT^4$

Using time dependant $G$ from equation (7) in equation (10) we get:

$$\frac{\dot{T}}{T} = \left(\frac{8\pi G_0 a}{3c^2}\right)^{1/2}\left(1 - \frac{T_{Pl}^2}{T^2}\right)^{1/2} T^2 \qquad \ldots (12)$$

Solving the above quadratic equation in $T$, we get the usual expression for time dependence of temperature as:

$$T = \left(\frac{3c^2}{32\pi G_0 a}\right)^{1/4} \frac{1}{t^{1/2}} \qquad \ldots (13)$$

This implies that the effects of temperature dependence of $G$ as given in equation (7) are too weak to be detected.

Another possible way in which the singular state may be avoided can be seen from modifying the Einstein-Hilbert action. In general relativity, the Einstein-Hilbert action yields the Einstein's field equations when varied to obtain equations of motion for the space-time metric.

The action $S[g]$ which gives rise to the vacuum Einstein equations is given by the following integral of the Lagrangian:

$$S[g] = \int \kappa R \sqrt{-g}\, d^4 x \qquad \ldots (14)$$

Where, $g = |g_{ab}|$ is the determinant of a space-time Lorentz metric, $R$ is the Ricci scalar, $\kappa = c^2/16\pi G$ is constant, the Lagrangian being $R\sqrt{-g}$, and the integral is taken over a region of space-time. The Einstein equations in the presence of matter are given by adding the Lagrangian for the matter into the integral.

The field equation is then obtained as:

$$R_{\mu\nu} - \frac{1}{2} g_{\mu\nu} R = \frac{8\pi G}{c^4} T_{\mu\nu} \qquad \ldots (15)$$

The Lagrangian need not necessarily be of the form $R\sqrt{-g}$. Instead the Lagrangian could be a function of the Ricci scalar, with the form $f(R)\sqrt{-g}$. This yields a field equation given by:[11, 12]

$$f'(R) R_{\mu\nu} - \frac{1}{2} g_{\mu\nu} f(R) = \frac{8\pi G}{c^4} T_{\mu\nu} \qquad \ldots (16)$$

Where $f(R)$ can have the form given by:[9, 10, 13]

$$f(R) = \frac{R}{\left(1 - R/R_{max}\right)} = \frac{R}{\left(1 - L_{Pl}^2 R\right)} \qquad \ldots (17)$$

Where $L_{Pl}^2 = \dfrac{1}{R_{max}}$ gives the maximal curvature

The motivation for choosing the form as given by equation (16) is by analogy with the Born-Infeld modification of Maxwell's electrodynamics.

Born and Infeld[14], tried to avoid the self energy divergence of a point charge in classical electrodynamics by incorporating (introducing) a maximal field strength $E_{max}$ and modifying the classical electrodynamic action as:

$$\alpha_E = \frac{E_{max}^2}{4\pi} \left[ 1 - \sqrt{1 + \left(\frac{F_{\mu\nu} F^{\mu\nu}}{2E_{max}}\right)^2} \right] \qquad \ldots (18)$$

In the limit of fields $E \ll E_{max}$, it reduces to the usual $\dfrac{1}{16\pi} F_{\mu\nu} F^{\mu\nu}$. The self energy calculated with the modified theory gives a finite value for the self energy and a minimal length scale as $r_{min}^2 = \dfrac{e}{E_{max}}$, $e$ being the electron charge.

Here the curvature is the equivalent to the field strength, and so the maximal curvature $R_{max} = 1/L_{Pl}^2$ plays the role of $E_{max}$ in the Born-Infeld theory.

Using this in equation (15), we get the field equation as:

$$\frac{R_{\mu\nu}}{\left(1 - L_{Pl}^2 R\right)^2} - \frac{2R_{\mu\nu}}{\left(1 - L_{Pl}^2 R\right)} = \frac{8\pi G}{c^4} T_{\mu\nu} \qquad \ldots (19)$$

On rearranging the terms we get:

$$R_{\mu\nu} - 2R_{\mu\nu}\left(1 - L_{Pl}^2 R\right) = \frac{8\pi G}{c^4} T_{\mu\nu} \left(1 - L_{Pl}^2 R\right)^2 \qquad \ldots (20)$$

As $R \to \dfrac{1}{L_{Pl}^2}$, the above equation reduces to:

$$R_{\mu\nu} = 0 \qquad \ldots (21)$$

That is the curvature vanishes, hence avoiding a singular state.

According to the equivalence principle, an accelerated frame is analogous to a gravitational field. Hence a maximal acceleration can imply a maximal field. The acceleration is given by the Riemann tensor as:

$$a^{\mu} = R^{\mu}_{\alpha\beta\gamma} \dfrac{dx^{\alpha}}{ds} \dfrac{dx^{\beta}}{ds} n^{\gamma} \qquad \ldots (22)$$

The maximal acceleration can be written as:

$$a = \dfrac{c}{t_{Pl}} = \dfrac{c}{(\hbar G/c^5)^{1/2}} \sim 10^{53}\, cm/s^2 \qquad \ldots (23)$$

Where, $t_{Pl} = (\hbar G/c^5)^{1/2} \sim 10^{-43}\, s$ is the Planck time.

This maximal acceleration can imply a minimum radius according to the relation:
(This is analogous to what we have in Born-Infeld case, that is $r_{min}^2 = e/E_{max}$)

$$r_{min}^2 = \dfrac{Gm}{a_{max}} = \dfrac{G^{3/2} m \hbar^{1/2}}{c^{7/2}} \qquad \ldots (24)$$

For the mass $m = m_{Pl} = (\hbar c/G)^{1/2}$, we have the above relation reducing to:

$$r_{min}^2 \approx \dfrac{1}{\Lambda_{Pl}} = 10^{-66}\, cm \qquad \ldots (25)$$

For the universe with the number density given by $N \approx 10^{90}$, the minimum size becomes:

$$r_{min} = N^{1/3} \times 10^{-33}\, cm = 10^{-3}\, cm \qquad \ldots (26)$$

As gravitational field around a black hole can give rise to temperature as suggested by Hawking[15], even an accelerating observer can detect a black body temperature given by the Unruh-Davis relation as:

$$T = \dfrac{\hbar a}{k_B c} \qquad \ldots (27)$$

For the above maximal acceleration the temperature corresponds to Planck temperature. The effect of this temperature arising from acceleration is being studied in accelerators and ion beams, where the effects are very much smaller than the maximal case.

This paper is a follow up of reference [1], where the present authors had discussed a possible scenario where inclusion of mutual gravitational interactions among the collapsing particles can avert a singularity. In this paper we have discussed the dynamics involved with the expanding universe from the Planck time to the present. We have also discussed other possible cases that can lead to a non-singular state, including temperature dependent gravitational constant and modified Einstein-Hilbert action, incorporating a maximal curvature (field strength).